# Dot-Science Top Level Domain: academic websites or dumpsites?[1]

**Enrique Orduña-Malea**

Universitat Politècnica de València

✉ Enrique Orduña-Malea
enorma@upv.es

**Abstract**:

Dot-science was launched in 2015 as a new academic top-level domain (TLD) aimed to provide 'a dedicated, easily accessible location for global Internet users with an interest in science'. The main objective of this work is to find out the general scholarly usage of this top-level domain. In particular, the following three questions are pursued: usage (number of web domains registered with the dot-science), purpose (main function and category of websites linked to these web domains), and impact (websites' visibility and authority). To do this, 13,900 domain names were gathered through *ICANN*'s *Domain Name Registration Data Lookup* database. Each web domain was subsequently categorized, and data on web impact were obtained from *Majestic*'s API. Based on the results obtained, it is concluded that the dot-science top-level domain is scarcely adopted by the academic community, and mainly used by registrar companies for reselling purposes (35.5% of all web domains were parked). Websites receiving the highest number of backlinks were generally related to non-academic websites applying intensive link building practices and offering leisure or even fraudulent contents. *Majestic*'s *Trust Flow* metric has been proved an effective method to filter reputable academic websites. As regards primary academic-related dot-science web domain categories, 1,175 (8.5% of all web domains registered) were found, mainly personal academic websites (342 web domains), blogs (261) and research groups (133). All dubious content reveals bad practices on the Web, where the tag 'science' is fundamentally used as a mechanism to deceive search engine algorithms.

# 1. Introduction

The *Internet Corporation for Assigned Names and Numbers* (*ICANN*) has dramatically increased the number of top-level domains (TLDs) since its inception. There are over 1,584 TLDs as of 12 February 2019.[2] Last TLDs delegated correspond to dot-gay (9 August 2019) and dot-cpa (20 September 2019).[3] Among all TLDs currently available we can distinguish a small number of domain names dedicated to academic activities.

The principal academic TLDs are oriented to organizations. This way, the first academic TLD was dot-edu (a sponsored TLD currently restricted to institutions located in the U.S., legally organized in the U.S., or recognized by a U.S. state, territorial, or federal agency). Later, dot-academy (2013), dot-college and dot-university (2014), and dot-mba (2015) were launched as an option for organizations that do not meet dot-edu's criteria, such as non-accredited institutions and institutions placed outside the United States.

An alternative option for academic websites oriented to organizations is the use of a second-level domain, for example, dot-edu. This option is currently followed by Australia (edu.au), Singapore (edu.sg), China (edu.cn), Malaysia (edu.my), Taiwan (edu.tw), Hong Kong (edu.hk), Poland (edu.pl), Bangladesh (edu.bd), and Pakistan (edu.pk). Additionally, we can find academic TLDs necessarily not oriented towards organizations, such as dot-education (December 2013), dot-science (November 2014), dot-courses (February 2015), dot-scholarships (April 2015) or dot-data (December 2016).

Dot-science stands out as a generic top-level domain (gTLD) officially aimed to provide 'a dedicated, easily accessible location for global Internet users with an interest in science' and becoming 'a source of information and services for online science communities'.[4] Unlike other academic domain names, dot-science is necessarily not related to academic institutions, but with scientific contents.

This gTLD was delegated to the root zone of the Domain Name System (DNS) by 15 November 2014 and is available since 24 February 2015. The applicant for this gTLD (referred to as the domain name registry) was *Famous Four Media*, a company headquartered in Gibraltar, which applied for a total of 61 gTLD.[5] The company has set up a separate Limited Company to act as the applicant for each individual application. *Dot Science Limited* was the applicant for dot-science. *Famous Four Media* also partnered with *Neustar Company* for the backend registry and technical requirements (back-end service provider).[6]

*Famous Four Media* was subsequently contracted by *Domain Venture Partners* (DVP) to manage 16 gTLDs. After a legal battle between these companies, *Global Registry Services Limited* (a company that belongs to DVP) started to manage the dot-science gTLD in 2018, by court-appointed administrator Edgar Lavarello, a Gibraltar-based accountant at *PricewaterhouseCoopers*.[7-8]

---

[2] https://www.iana.org/domains/root/db
[3] https://newgtlds.icann.org/en/program-status/delegated-strings
[4] http://nic.science/why.html
[5] https://domaintyper.com/new-gTLD/applicant/Famous-Four-Media
[6] https://www.iana.org/domains/root/db/science.html
[7] http://domainincite.com/23284-famous-four-is-dead-new-registry-promises-spam-crackdown

Despite the potential importance of the dot-science TLD for online scientific contents dissemination, no systematic studies on this academic web space have been conducted to date, beyond unpublished brief technical reports (*W3 Techs* n.d.) or professional blog posts (Larsen 2015). Consequently, there is no data about what content is available under the dot-science label, or what impact or influence this content may have. Issues that can be partially explored applying webometric techniques.

The main goal of this work is to determine the scholarly use of the dot-science gTLD. To do this, a webometric approach to the whole dot-science domain is applied in order to determine its intensity of use, purpose of use, and web visibility.

To achieve the goals established, the following research questions are set up:

RQ1: How much is the dot-science gTLD used by the community (in terms of users' demand)?
RQ2: What is the dot-science gTLD used for (in terms of users' purposes and types of websites created)?
RQ3: What impact has dot-science web domains reached (in terms of web visibility and reputation achieved)?

## 2. Research Background

Webometrics has been defined as the study of web-based contents with primarily quantitative methods for social research goals, using techniques that are not specific to one field of study (Thelwall 2009), and heavily framed within Informetrics (Bjorneborn and Ingwersen 2004). Webometrics has been also referred to as informetric analysis of the Web (Bar-Ilan 2001).

Even though pioneering webometric studies started in the mid-nineties of the last twentieth century, the official baptize came with the seminal work by Almind and Ingwersen (1997), who coined the name 'webometrics' (Thelwall 2012a), and the launch in 1997 of a journal devoted to the field (*Cybermetrics: International Journal of Scientometrics, Informetrics and Bibliometrics*), by Isidro Aguillo as editor in chief.

Despite Webometrics refers to the whole analysis of the web phenomena, covering not only contents but also web services and technologies regardless the topic (Bjorneborn and Ingwersen 2004; Orduna-Malea and Aguillo 2015), the application of this field was primarily focused to the Science of Science (Fortunato et al. 2018), with the purpose of understanding the scholarly use of the Web (Barjak 2006) and evaluate its potential to expand the notion of academic impact and research evaluation (Thelwall and Kousha 2015). The similarity between Webometric and Bibliometric indicators (both with common strands with Informetrics) as well as the strong and solid structure of the scientific community facilitated the application of Webometrics to the broad fields of Scientometrics, Bibliometrics, and research evaluation (namely, academic Webometrics).

---

[8] http://domainincite.com/23302-i-was-wrong-famous-four-bosses-were-kicked-out

Academic Webometrics evolved in two complementary directions. On one side of the spectrum, new methods and models (Bjorneborn and Ingwersen 2004; Thelwall and Sud 2011; Thelwall 2012a), techniques (Park and Thelwall 2003; Thelwall 2004; 2006; Thelwall et al. 2012) and indicators (Ingwersen 1998; Thelwall and Kousha 2003; 2012; 2015; Aguillo et al. 2006; 2010; Bollen and Van de Sompel 2008; Kousha et al. 2010) were designed and studied.

On the other side of the spectrum, all this new knowledge acquired was directly applied to the analysis of authors, organizations and products related to the scientific endeavor, such as personal websites (Barjak et al. 2007; Más-Bleda and Aguillo 2013; Más-Bleda et al. 2014), universities (Payne and Thelwall 2007; Aguillo et al. 2008; Ortega and Aguillo 2009; Vaughan and Romero-Frías 2014), university units and services (Li et al. 2005a; 2005b; Orduna-Malea 2013), articles (Bollen and Sompel 2008), journals (Vaughan and Thelwall 2003; Bollen and Sompel 2008; Bollen et al. 2009; Thelwall 2012b), repositories (Aguillo et al. 2010; Orduna-Malea and Delgado López-Cózar 2015), hospitals (Utrilla-Ramírez et al. 2011) or scientific parks (Minguillo and Thelwall 2012).

The huge dependency on commercial search engine functionalities, the variability of data, and the ease of data manipulation limited the development of the field (Thelwall 2010), which was overcome by the rise of Altmetrics (Warren et al. 2017). However, the use of webometric methods is still advisable as a supplementary source of information related to academic/scientific impact and research communication.

Traditional webometric techniques, applied either to organizations or products related to Science and scientific communication, were performed regardless the TLD used by the websites. However, the appearance of top-level domains exclusively oriented to disseminate academic and scientific contents, such as dot-science, opens new possibilities to better understand the scholarly use of the Web.

A TLD dedicated to scientific contents may attract wide range academic actors (journals, authors, organizations, companies, etc.) which may be interested to use the dot-science gTLD for branding purposes and online reputation. However, the analysis of all the web domains registered under the umbrella of one specific gTLD (in this case, dot-science) has not been carried out to date. This study aims to fills this gap in the field of academic Webometrics.

## 3. Method

Data about dot-science usage were obtained from the monthly registry reports (transaction reports and activity reports) provided by ICANN.[9] All data was extracted in csv file for statistical analysis.

All web domains registered under the dot-science TLD were directly extracted from ICAAN's *whois* directory.[10] A total of 13,900 web domains were gathered as of 8 February 2020.

---

[9] https://www.icann.org/resources/pages/science-2015-03-01-en
[10] https://whois.nic.science

As Google does not offer features for massive link analysis, alternative tools are needed. The search engine optimization (SEO) industry provides tools with specific features to analyze the number of hyperlinks that web domains receive, among other web metrics. Among link analytics tools, Majestic,[11] Ahrefs,[12] Link Explorer[13] and SEMRush[14] stand out. While all these tools offer similar features and metrics, Majestic exhibits some advantages for webometric studies. First, Majestic is centered on backlinks offering a wider range of tailored link-based metrics. Specifically, Trust Flow and Topical Trust Flow (see Table 1) are metrics not offered by any other tool. Second, Majestic is one of the most comprehensive sources of backlinks data on the Web, declaring 2,481 billion unique URLs indexed in its historic database (coverage from 2015 to 2020) and 450 billion unique URLs indexed in its fresh database (last five months), as of 30 Nov 2020.

Taking these considerations into account, data about the web impact achieved by dot-science web domains were gathered from *Majestic* API[15] through the fresh database. This database was selected because it covers recent link-based data, excluding deleted URLs as much as possible, providing thus a more updated overview of dot-science web domains. JSON files were obtained, cleaned through *OpenRefine*, and finally exported to a spreadsheet file for statistical analysis. All metrics gathered are summarized in Table 1. This process was carried out by 17 February 2020.

**Table 1**
Summary of metrics used: source and scope

| Metric | Source | Scope |
|---|---|---|
| DNS queries | ICANN | Number of requests sent from a user's computer (DNS client) to a DNS server. It is use as a proxy of a demand for information. |
| *Whois* queries | ICANN | Number of requests sent from a *whois* database referencing a top-level domain. It is use as a proxy of a demand for information. |
| URLs indexed | Majestic | The number of URLs created within a web domain and indexed by one search engine. This parameter is used as a proxy of website volume, in terms of contents. |
| External backlinks | Majestic | The number of hyperlinks received by a web domain from websites registered in other web domains. This parameter is used as a proxy of website visibility and interest. |
| Educational external backlinks | Majestic | The number of hyperlinks received by a web domain from dot-edu websites registered in other web domains. |
| Internal outlinks | Majestic | The number of hyperlinks generated by webpages under a web domain, and targeted to other webpages hosted in the same web domain. These hyperlinks are used for navigational purposes. |
| External outlinks | Majestic | The number of hyperlinks generated by webpages under a web domain, and targeted to other webpages hosted in different web domains. These hyperlinks are mainly used for informative purposes. |
| Referral domains | Majestic | The number of web domains from which a web domain receives at least one external backlink. This parameter is used as a proxy of website visibility and interest. |
| Educational referral domains | Majestic | The number of dot-edu web domains from which a web domain receives at least one external backlink. |
| Trust Flow | Majestic | Score on a scale between 0-100 achieved by one web domain, based on the number of hyperlinks it receives from trusted seed sites. This parameter is used as a proxy of website trustworthy. |
| Topical Trust Flow | Majestic | Score on a scale between 0-100 achieved by one web domain, which shows the relative influence of this web domain in any given topic or category. The category of one web domain is based on the category of |

---

[11] https://majestic.com
[12] https://ahrefs.com
[13] https://moz.com/link-explorer
[14] https://www.semrush.com
[15] https://developer-support.majestic.com/api

| | | |
|---|---|---|
| | | those web domains linking to it. This parameter is used as a proxy of website authoritative on a certain topic. |
| Citation Flow | Majestic | Score on a scale between 0-100 achieved by one web domain, based on the number of hyperlinks it receives. This parameter is used as a proxy of website influence. |

A functional taxonomy was built in order to classify all web domains previously gathered. The classification provided by Halvorson et al. (2012) was used as a starting point, being subsequently expanded and adapted to fulfill the requirements of this work.

The taxonomy is based on four main functions (primary, defective, defensive, and parked web domains), each of which holds different categories and subcategories, based on the purposes of the website:

- *Primary web domain*: the registrant uses the web domain to identify itself, its service, or online resources.
- *Defensive web domain*: the registrant uses the web domain to defend an alternative web domain. As a consequence, a direct redirect from dot-science web domain to another web domain is found.
- *Parked web domains*: the registrant has not published web-related content in the host where the dot-science web domain has been registered. As a consequence, a default page is found. This page can be used for different purposes (links to promote other websites, reselling information, registrar company information, default webpage template, etc.). The registrant uses to be the registrar company.
- *Defective web domains*: no website is served to the final user. This can be due to the existence of technical errors or expired web domains. As a consequence, there is no way to determine whether the website is primary or parked.

The classification process was performed in a three-stage process. First, all websites were accessed manually, and a general and preliminary classification was established (primary, defensive, parked, defective). Second, subcategories were annotated following a bottom-up approach (inductive method that consists of observing and categorizing each website, finding patterns, and grouping websites by these common patterns). Finally, category and subcategory names were standardized and grouped (Table 2). Third, all web domains were accessed again checking their general function and including the corresponding category and subcategory. This process was carried out throughout March 2020.

**Table 2**
Taxonomy of dot-science top-level web domains according to their functionality and purpose

| Function | Category | Subcategory |
|---|---|---|
| Primary | Closed | Maintenance; Not currently available; Under construction |
| | Company | Academy; Clinic; Product; Service; Software |
| | General content | Adult contents; Animations; Application; Casinos; Code; Comics; Downloads; Fora; Games; Governmental information; Humor; Images collection; informal shopping; informational resources; Interactive tools; Job information; mail services; maps; Mirror sites; Music; Quiz; Radio tracks; Radio tunes; RSS Feeds; Social bookmarks; Social News Managers; Spreadsheets; Touristic information; Twitter timelines; Videos collection; Videogames; Videogame content; Videos; Wedding days; Wikis. |
| | Academic content | Academic videos; Books; Databases; Directories; Educational resources; Information resources; Journals; Publication supplements; Surveys; Theses. |
| | Empty | Incomplete Website; Index of Page; Text |
| | Event | |
| | Organization | Alumni association; Scientific association; Club; Consortium; Cultural association; Foundation; Health institution; High School; Institute; Library; Museum; National Research Foundation; NGO; Research Center; Research Community; Research Council; Research Federation; Research group/lab; Research institute; Research Society; School; Scientific Academy; Social group; Standardization institution; University; University Department; University Master Degree; Working group; Palace |
| | Research Platform | |
| | Research Projects | |
| | Personal Website | |
| | Portal | |
| | Private | Intranets |
| | Thematic blog | Academic-related; Non-academic related |
| | Thematic website | Academic-related; Non-academic related |
| | Flat Page | Animated text; Brief information piece; Background image; Button; Hidden text; image; Logo; Text; Video; |
| | Bait Page | |
| Defective | Dropped | Blocked; Deleted; Disabled; Expired; On hold; Suspended |
| | Error | |
| Defensive | | |
| Parked | Registrar Page | |
| | Related Links Page | |
| | Text Page | |
| | Web Default Page | |
| | Web Customized Page | |
| | Web Server Default Page | |

To check the accuracy of the classification process, an external researcher with expertise in web science was asked to carry out an inter-coder reliability test through a random sample of 200 web domains. The percentage of agreement achieved was low (61%). However, after the identification of few systematic errors (few subcategories were confused and assigned wrongly) and a subsequent training, the concordance achieved 83%, which is considered acceptable.

# 4. Results

## 4.1. Science top-level web domain activity

The first web domain registered with the dot-science gTLD dates from November 2014, that is, nine months after the official launch of this academic gTLD. Since then, the number of registered web domains rapidly grew to a maximum of 349,860 in May 2016. From that moment on the number of dot-science web domains decreased to 14,078 (December 2019). Two specific moments (August-September 2016, and especially July 2017) determined this evolution.

As we can observe in Figure 1, web domains were managed from few registrars. *Paradise Registrars* held a great percentage of dot-science web domains (80.9%–94.9%) from 2015 to 2017. This company finished its domain activity in March 2017, and passed the baton to the company *Alpnames*. However, this company was de-accredited by ICAAN in March 2019, then ceasing all its domain activity.[16] Recently, *NameCheap* registrar is responsible for the majority of dot-science web domains (79.6% of all dot-science web domains were managed by this company as of June 2019). However, the number of dot-science web domains registered by this company started to drop at the end of 2019.

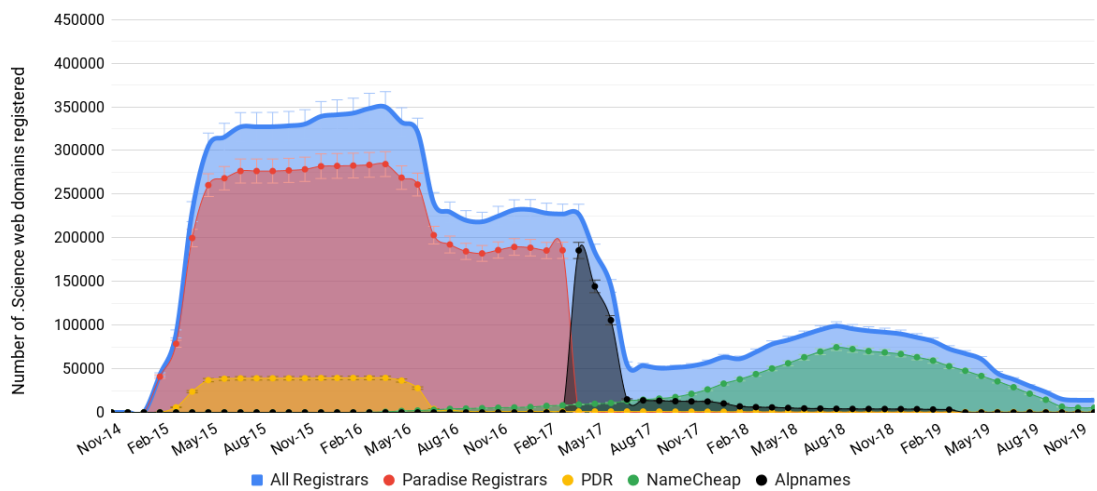

**Figure 1**
Number of dot-science top-level domains registered broken down by year

It is noteworthy to mention the country location of registrars. United States (30.6%) and China (20.1%) stand out as the places with most dot-science gTLDs registered as of February 2020, mainly due to *NameCheaps* and *GoDaddy* in the case of United States, and *Alibaba Cloud Computing* in the case of China. France (10.1%) and Panama (7.2%) appear in third and fourth place, respectively.

The evolution of DNS queries from November 2014 to December 2019 is offered in Figure 2, where a decrease from March 2016 to May 2018 is observed. This period corresponds to the time when *Paradise Registrars* experienced a first important drop and *NameCheap* started increasing the number of dot-science web domains registered (Figure 2). The number of *whois* queries is offered in Figure 3. In this particular case, the number of queries from April 2017 onwards is practically non-existent.

---

[16] https://www.domainpulse.com/2019/03/15/alpnames-is-no-more-as-icann-terminates-registrar-days-after-going-offline

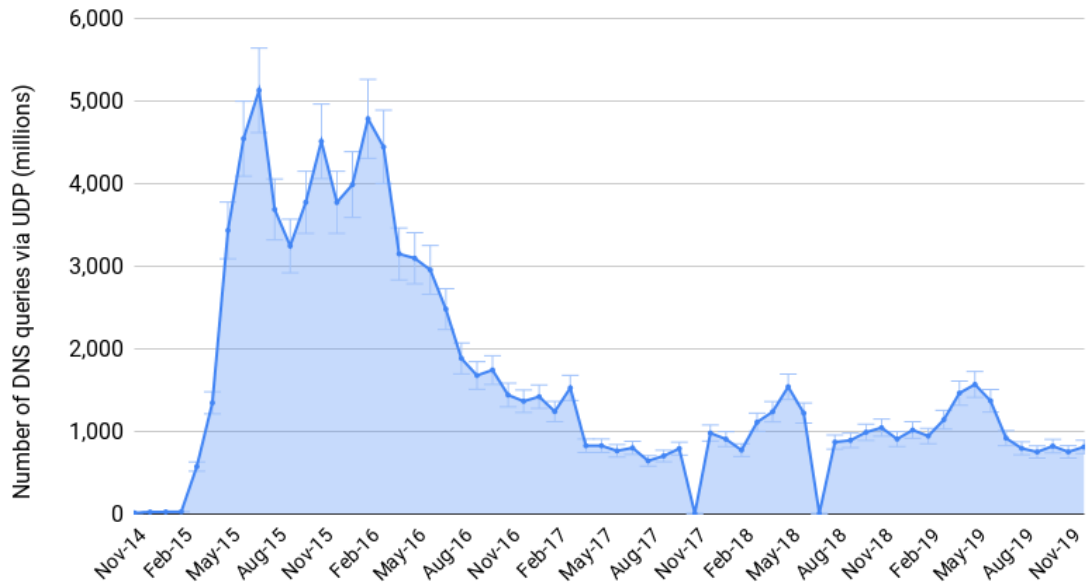

**Figure 2**
Number of DNS queries through UDP protocol performed over dot-science gTLDs broken down by year

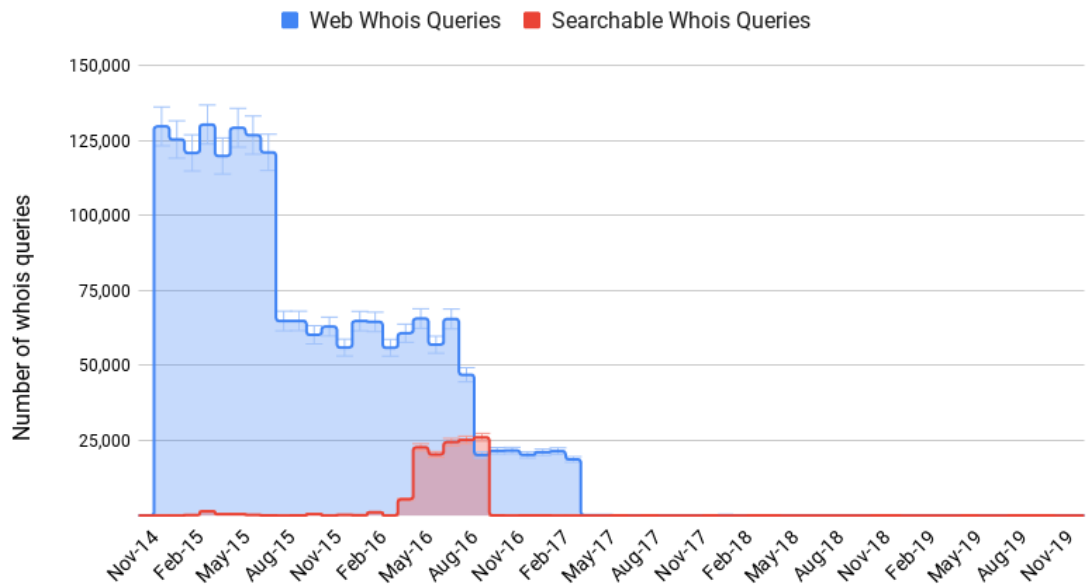

**Figure 3**
Number of *whois* queries (web-based and searchable) performed over dot-science gTLDs broken down by year.

### 4.2. Web category

38.7% of all web domains analyzed (5,381) were defective, while 35.5% (4,932) were parked, 16.0% (2,222) were primary, and 9.8% (1,365) were defensive. The frequency of the main categories related to each web domain function is offered in Table 3.

**Table 3**
Frequency of categories of websites registered with the dot-science gTLD

| DEFECTIVE | N | % | PARKED | N | % | DEFENSIVE | N | % | PRIMARY | N | % |
|---|---|---|---|---|---|---|---|---|---|---|---|
| **Error** | 5,173 | 96.1 | Text page | 1,879 | 38.1 | Redirected | 1,365 | 9.8 | **Scholarly categories** | **793** | **36** |
| **Dropped** | 208 | 3.9 | Registrar page | 1,620 | 32.8 | | | | **Non-Scholarly categories** | **955** | **43** |
| TOTAL | **5,381** | **100** | Related links page | 1,125 | 22.8 | | | | Empty | 323 | 14.5 |
| | | | Server default page | 230 | 4.7 | | | | Company | 291 | 13.1 |
| | | | Web default page | 78 | 1.6 | | | | Flat page | 159 | 7.2 |
| | | | **TOTAL** | **4,932** | **100** | | | | Private (intranets) | 85 | 3.8 |
| | | | | | | | | | Closed | 64 | 2.9 |
| | | | | | | | | | Bait page | 33 | 1.5 |
| | | | | | | | | | **Miscellany** | **474** | **21** |
| | | | | | | | | | Thematic blog | 261 | 11.7 |
| | | | | | | | | | General content | 142 | 6.4 |
| | | | | | | | | | Thematic website | 71 | 3.2 |
| | | | | | | | | | **TOTAL** | **2,222** | **100** |
| **38.7%** | | | **35.5%** | | | **9.8%** | | | **16%** | | |

Defective web domains were mainly characterized by access errors. The 'IP Not Found' message was obtained from 3,117 web domains. Other common errors were: 'Connection Time Out' (504), 'Connection Refused' (356), 'Blank Page' (335), 'Bad Gateway' (217) or 'Page Not Found' (215). Otherwise, up to 208 web domains were actually accessed. However, a default message indicating web domain suspension (109 web domains), expiration (84) or any other similar circumstance was obtained.

Parked web domains were found to be mainly performed by default text pages without images or links to other websites (1,879 web domains). Generic pages including information related to the services offered by the registrar and generic pages containing lists of links targeting to other websites were found to be also a common practice (1,620 and 1,125 web domains, respectively). Among the parked web domains we can find names of famous universities (such as 'dukeuniversity.science', 'uchicago.science' or 'ucla.science') or journals (e.g., 'physicalreviewapplied.science' or 'physicalreviewletters.science'). Other parked web domains include academic-related words to be offered at a cost (for example, 'turingtest.science' is offered at a sale price of 798 USD).

Defensive web domains constitute a small but significant function category. A total of 1,365 web domains redirected to 889 different web domains (Table 4), out of which 73 were universities (especially from United States and Germany). The most frequently destination corresponded to 'mayafreebird.com', an empty website with no contents publicly published, followed by 'dan.com', a registrar company website. Other registrars ('findresultsonline.com', 'decent.domains', and 'uniregistry.com') also achieved top positions. Other destinations to which dot-science web domains frequently redirected to were 'forbid4f.info', a company selling miticide insecticide, and 'malvernpanalytical.com', a scientific instrumentation company provider.

Other defensive strategies were found for well-established companies and services (e.g., 'facebook.science', 'instagram.science', 'amazon.science', imdb.science', 'microsoft.science' or 'baidu.science'), including academic-related databases (e.g., 'scopus.science', 'clarivate.science', 'altmetric.science'), whose dot-science web domains redirected to their official websites.

**Table 4**
Most frequently destinations to which dot-science web domains redirect to

| Web domains | N | Top-level Domains | N |
|---|---|---|---|
| mayafreebird.com | 61 | .com | 739 |
| dan.com | 43 | .org | 96 |
| google.com | 24 | .de | 65 |
| malvernpanalytical.com | 21 | .net | 35 |
| linkedin.com | 21 | .science | 33 |
| findresultsonline.com | 17 | .edu | 33 |
| decent.domains | 17 | .info | 31 |
| uniregistry.com | 13 | .io | 24 |
| forbid4f.info | 13 | .fr | 21 |
| wordpress.com | 12 | .domains | 19 |

Primary web domains amount to 2,222. This category was subsequently divided into scholarly-related categories (36%), non-scholarly related categories (43%), and a 'general category', containing websites that might or might not be oriented to scholarly activities (21%).

Non-scholarly categories included both incomplete and complete websites. Among the incomplete websites we can find 323 empty websites (generally a CMS with a default theme template installed, but no contents published), 64 closed (including an image or text announcing 'under construction', 'maintenance' or 'coming soon'), and 33 bait pages (webpages without contents but just one or more links to other places, acting as defensive websites but without an automated redirect).

Among the non-scholarly complete websites, private companies stand out (291 web domains). Within this category, we found 216 service-based companies (out of which 29 offered software services), product-based companies (67 web domains), and private clinics (8 web domains).

As regards the miscellany category, the presence of thematic blogs is quite remarkable (261 websites). However, around 30% of these websites cannot be considered academic-related blogs. This problem was even greater with those web domains typified under the 'General content' category. Despite this category included interesting content such as software applications (26) or interactive tools (11), it also covered video reviews about sale products (23), pornographic content (8), information related to casinos (6), websites dedicated to massive content download (5), videogames (5), social news managers (5) or even websites covering wedding days (2).

Please see section 4.4 for a detailed analysis of academic-related primary websites.

### 4.3. Web impact

**Website volume**

*Majestic* does not provide the number of URLs indexed for 4,797 web domains (34.5%), mainly defective (2,072) and parked (1,889). Primary web domains with low web traffic might also be uncovered by *Majestic* database.

Considering the remaining 9,103 web domains, the distribution of the number of URLs indexed is heavily skewed (Figure 4). 223 web domains showed at least 1,000 URLs indexed, while 548 web domains showed at least 100, and 5,367 web domains showed just one URL indexed.

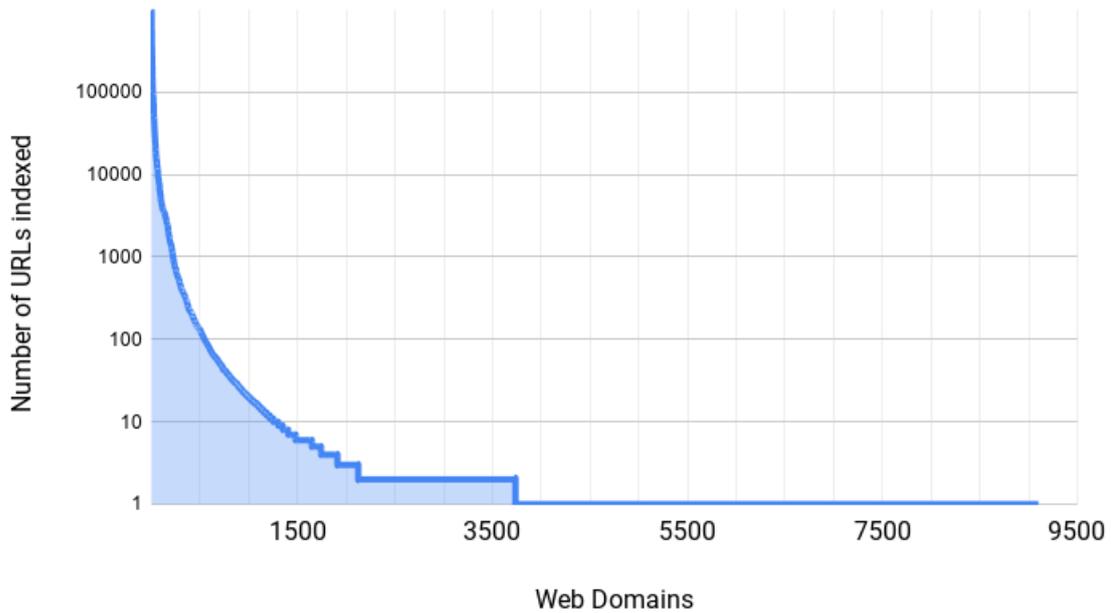

**Figure 4**
Distribution of the number of URLs indexed for dot-science top-level domains

Big web domains (those with at least 1,000 URLs indexed) are mainly integrated by primary web domains (69%). However, big defensive (7.4%) and parked web domains (6%) were also found, which demonstrate the existence of internal URLs (with or without content publicly published) that cannot be reached by browsing from the homepage.

Table 5 includes information related to all web domains with at least 100,000 URLs indexed. We can observe the existence of web domains dedicated to offer academic-related content, such as databases, *GitHub* projects directories or research organizations. The biggest website corresponds to an incomplete website (a consulting company with a website's backend devoid of content), followed by a defensive website which redirects to a company. It is also worth to note the existence of pornography and videogame content (list of advanced commands to play *Minecraft* videogame) amongst the most voluminous web domains.

**Table 5**
Dot-science web domains with the highest number of indexed URLs (*Majestic*)

| Web domain | Indexed URLs | Function | Category | Subcategory |
|---|---|---|---|---|
| aleryon.science | 998,851 | Primary | Empty | Incomplete website |
| rttv.science | 958,059 | Defensive | No available | No available |
| communitydata.science | 946,840 | Primary | Organization | Research community |
| moonbutt.science | 637,766 | Primary | Academic content | Directory |
| proceedings.science | 436,569 | Primary | Academic content | Database |
| alephzero.science | 369,911 | Primary | Academic content | Directory |
| incest.science | 341,315 | Primary | General content | Adult (explicit videos) |
| blaschke.science | 294,959 | Parked | Web default page | No available |
| furry.science | 259,546 | Primary | General content | Adult (videogame) |
| downloadsach.science | 242,230 | Primary | General content | Downloads |
| minecraftcommand.science | 175,920 | Primary | General content | Videogame content |
| goodwolf.science | 155,742 | Primary | Empty | Incomplete website |
| fias.science | 139,795 | Primary | Organization | Research institute |
| brainstem.science | 138,526 | Primary | Organization | Research group |
| lehrbuch.science | 101,904 | Primary | Academic Content | Database |
| nes.science | 100,056 | Primary | Thematic Website | Technical information |

**Website visibility**

The distribution of the number of external backlinks received by the web domains registered with the dot-science gTLD is also heavily skewed. 6,347 (69.7%) out of the 9,103 web domains indexed by *Majestic* do not receive external backlinks, while 1,398 receive just one, 343 receive at least 100, and only 157 receive at least 1,000.

The web domain with the highest number of external backlinks received corresponds to a company selling Viagra online (1,819,506 links from 7,763 different web domains). It is worth to note that this web domain receives 2,484 links from 65 different dot-edu web domains and 4,396 links from 17 different dot-gov web domains (Table 6).

The web domains which receive links from a larger number of different web domains (referral domains) are found to be all of them standalone webpages created by *MediaWiki* open source wiki engine, with the default title 'Main Page'. Each page includes one outlink targeted to a product-based company. This is a black SEO professional practice known as 'tier', used to inflate a website's reputation.

**Table 6**
Dot-cience web domains with the highest number of referral domains (*Majestic*)

| Web domain | Domains ALL | Domains EDU | Domains GOV | Links ALL | Links EDU | Links GOV | Content |
|---|---|---|---|---|---|---|---|
| overthecounterviagra.science | **7,763** | 65 | 17 | 1,819,506 | 2,484 | 4,396 | Product |
| elearnportal.science | **2,596** | 19 | 3 | 110,500 | 162 | 23 | Wiki |
| scientific-programs.science | **2,539** | 20 | 4 | 94,160 | 477 | 32 | Wiki |
| sciencewiki.science | **2,523** | 29 | 6 | 105,889 | 202 | 202 | Wiki |
| yogicentral.science | **2,512** | 27 | 5 | 120,818 | 219 | 17 | Wiki |
| mozillabd.science | **2,467** | 22 | 4 | 100,588 | 181 | 30 | Wiki |
| nerdgaming.science | **2,425** | 23 | 5 | 223,263 | 241 | 26 | Wiki |
| yogaasanas.science | **2,415** | 24 | 7 | 83,853 | 231 | 27 | Wiki |
| opensourcebridge.science | **2,413** | 15 | 4 | 110,028 | 243 | 225 | Wiki |
| wifidb.science | **2,406** | 22 | 5 | 109,890 | 340 | 19 | Wiki |
| spamdb.science | **2,303** | 20 | 4 | 92,608 | 214 | 21 | Wiki |
| ai-db.science | **2,299** | 20 | 5 | 104,306 | 471 | 38 | Wiki |
| securityholes.science | **2,252** | 23 | 6 | 92,455 | 296 | 36 | Wiki |
| morphomics.science | **2,247** | 24 | 4 | 85,914 | 232 | 29 | Wiki |
| pediascape.science | **2,236** | 15 | 4 | 89,418 | 119 | 19 | Wiki |
| phonographic.science | **2,026** | 19 | 3 | 84,469 | 185 | 19 | Wiki |

Another online behavior detected is the existence of web domains receiving great amounts of hyperlinks from few sources. The most extreme cases are included in Table 7. For example, the parked web domain 'essaywriting.science' receives 94,432 links from just one web domain. Also, a set of web domains offering product sales video galleries (all of them created with the same CMS template) receive a similar number of hyperlinks (around 220,000) from a similar number of referral domains (around 90).

**Table 7**
Dot-science web domains with large number of backlinks received from few referral domains (*Majestic*)

| Web domain | Domains | Backlinks | Function | Category | Subcategory |
|---|---|---|---|---|---|
| essaywriting.science | 1 | 94,432 | Parked | Related links page | |
| lieberman.science | 12 | 345,281 | Primary | Organization | Research group |
| sgcbuilding.science | 6 | 46,199 | Primary | General content | Videos collection |
| netias.science | 25 | 190,929 | Primary | Organization | Research community |
| goethean.science | 25 | 138,658 | Primary | Organization | Research community |
| humboltedu.science | 1 | 2,794 | Primary | Thematic blog | |
| proteogenix.science | 130 | 340,817 | Primary | Company | Product |
| sunglasses.science | 84 | 219,221 | Primary | General content | Videos collection |
| fooding.science | 85 | 218,169 | Primary | General content | Videos collection |
| snowboard.science | 86 | 218,744 | Primary | General content | Videos collection |
| surfboard.science | 86 | 218,133 | Primary | General content | Videos collection |
| stockmarkets.science | 87 | 219,506 | Primary | General content | Videos collection |
| jewellery.science | 87 | 218,235 | Primary | General content | Videos collection |
| printers.science | 88 | 219,772 | Primary | General content | Videos collection |
| cloudserver.science | 88 | 219,062 | Primary | General content | Videos collection |
| shaving.science | 89 | 219,760 | Primary | General content | Videos collection |
| golfgear.science | 89 | 219,735 | Primary | General content | Videos collection |
| bluejeans.science | 90 | 220,407 | Primary | General content | Videos collection |
| motorbike.science | 90 | 219,280 | Primary | General content | Videos collection |
| moneymaker.science | 90 | 218,341 | Primary | General content | Videos collection |

*Majestic* does not provide information on outlinks for 5,895 indexed web domains, while shows 3,129 web domains with at least one outlink, either internal or external, as well as 79 web domains with no outlinks.

If we focus on the external outlinks, a general low usage is observed (mean: 4.7 outlinks; median: 1 outlink). Only seven web domains had created at least 100 external outlinks. The web domain with the highest number of external outlinks corresponded to 'vww.science', which exhibited 1,890 external outlinks. However, this web domain has been recently removed (currently shows 'Error 404 Page Not Found').

**Website reputation**

Results evidence low *Citation Flow* values (mean: 1.57; median: 0). 74.3% out of the 9,103 web domains indexed by *Majestic* achieved a *Citation Flow* score equal to 0. A database offering access to conference proceedings ('proceedings.science') obtained the highest *Citation Flow* value (49), followed by a research center ('crest.science'; 43) and a research group ('laet.science'; 38).

The web domains with the highest *Trust Flow* values are offered in Table 8. As we can observe, this metric was capable of filtering primary web domains offering academic-related content. Notwithstanding, *Trust Flow* values are low (mean: 0.49; median: 0). As evidence of this, 93.5% out of the 9,103 web domains found by *Majestic* obtained a *Trust Flow* value equal to 0.

**Table 8**
Web domains with the highest *Trust Flow* values (*Majestic*)

| Web domain | Trust Flow | Citation Flow | Category | Subcategory |
|---|---|---|---|---|
| cerf.science | **41** | 36 | Organization | Research federation |
| council.science | **40** | 37 | Organization | Research council |
| bfh.science | **32** | 27 | Thematic website | Technical information |
| nouvelle-aquitaine.science | **28** | 34 | Organization | Research community |
| proteogenix.science | **28** | 34 | Company | Product |
| amazon.science | **27** | 34 | Thematic website | General scientific content |
| tvc-16.science | **27** | 22 | Thematic blog | |
| lindinglab.science | **26** | 29 | Organization | Research group |
| crest.science | **25** | 43 | Organization | Research center |
| family.science | **25** | 20 | Company | Service |

Only a small number of web domains (558) received a categorization through the *Topical Trust Flow* metric. The topics on which these web domains were most authoritative were quite diverse (Table 9), highlighting the presence of 'Reference/Education' (most authoritative topic for 63 web domains), and 'Science/Biology' (34). However, the *Trust Flow* associated to these topics was low. 'Society/Philosophy' (11.79) exhibited the highest average *Topical Trust Flow*.

**Table 9**
Topics in which the dot-science web domains are most frequently categorized, and their average *Topical Trust Flow* (*Majestic*)

| TOPIC | FREQUENCY | AVERAGE TOPICAL TRUST FLOW |
|---|---|---|
| Reference/Education | 63 | 6.98 |
| Science/Biology | 34 | 10.59 |
| Business | 30 | 2.13 |
| Science/Social Sciences | 23 | 8 |
| Regional/North America | 18 | 1.61 |
| Science/Math | 16 | 11 |
| Computers/Software/Operating Systems | 14 | 6.57 |
| Society/Philosophy | 14 | 11.79 |
| Computers/Programming/Languages | 13 | 8.08 |
| Science/Earth Sciences | 13 | 10.69 |
| Science/Technology | 13 | 8.31 |
| Regional/Europe | 12 | 9.83 |
| Computers/Internet/Web Design and Development | 11 | 9.73 |
| Science/Physics | 11 | 9.91 |
| Science/Astronomy | 10 | 5.8 |
| Society/People | 10 | 7.3 |

*Citation Flow* correlates moderately, though statistically significantly, with the *Trust Flow* (0.5; α > 0.01; p-value: < 0.0001). Figure 5 includes a scatter plot of these two metrics. As we can observe, a number of web domains obtained significantly higher *Citation Flow* scores than *Trust Flow* scores. These websites attracted a great quantity of external backlinks, but from low trusted web domains.

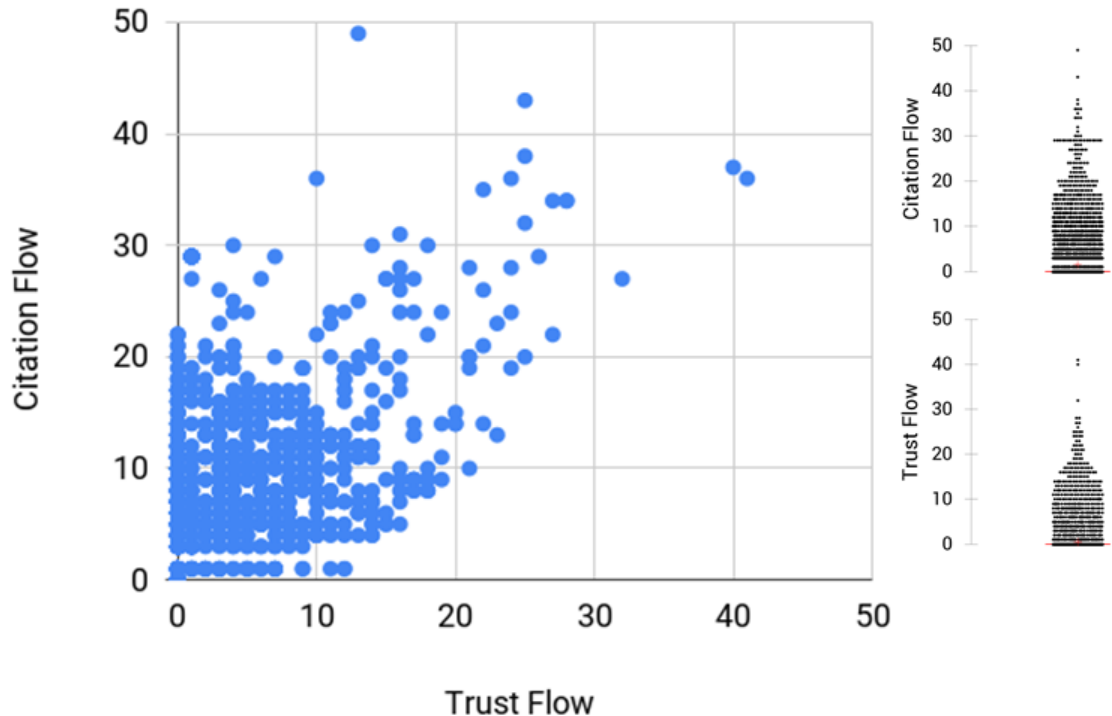

**Figure 5**
Distribution of *Trust Flow* and *Citation Flow* for dot-science gTLDs (n= 9,103)

### 4.4. Scholarly primary websites

The academic community has used dot-science gTLDs to build a wide variety of academic websites (51 different typologies related to academic activities, services, and products have been identified). The websites most frequently hosted in these primary scholarly web domain categories correspond to academic personal websites (324 web domains, mainly offering scholars' personal information, publications lists, and contact). Organizations constituted another important website category. It was mainly integrated by research groups and laboratories (133 web domains), research communities (37), research centers (27), scientific associations (13) and research societies (8). Finally, the 'Academic content' category comprised mainly databases (31), journals (22), directories (10), books (7) and information resources (5). Additionally, supplementary material to publications, such as detailed explanations for posters or articles published, were also found (17).

Table 10 includes all primary scholarly-related web categories, which shapes the academic usage of the dot-science gTLD (including general content related to scholarly activities). The URLs indexed average and web impact for each category are also available. Since some categories include few web domains, averages values were preferred instead of median values.

**Table 10**
Primary dot-science scholarly categories: content generation and impact

| Category | Subcategory | N | Avg UI | Avg RD | Avg EBL | Avg EOL | Avg TF | Avg CF |
|---|---|---|---|---|---|---|---|---|
| Scholarly content | Academic videos | 2 | 572 | 0 | 0 | 2 | 0 | 0 |
| | Books | 7 | 8 | 2 | 14 | 1 | 1 | 3 |
| | Catalog | 2 | 211 | 25 | 177 | 4 | 5 | 11 |
| | Databases | 31 | 23,197 | 26 | 15,129 | 3 | 3 | 9 |
| | Directories | 9 | 788 | 5 | 4,251 | 62 | 3 | 4 |
| | Educational resources | 4 | 6,969 | 2.5 | 70 | 1 | 5 | 6 |
| | Encyclopedias | 1 | 9,177 | 0 | 0 | NA | NA | NA |
| | Information resources | 5 | 91 | 12 | 37 | 6 | 4 | 6 |
| | Publication supplements | 17 | 435 | 5 | 39 | 2 | 4 | 5 |
| | Research survey | 3 | 2 | 1 | 0 | 1 | 0 | 5 |
| | Thesis | 1 | NA | 0 | 0 | NA | NA | NA |
| Events | | 26 | 6,430 | 5 | 24 | 5 | 4 | 6 |
| General content | Applications | 26 | 1,158 | 12 | 199 | 5 | 3 | 6 |
| | Code | 11 | 6,090 | 14 | 1,237 | 5 | 2 | 8 |
| | Governmental information | 2 | 2,079 | 6 | 884 | 2 | 0 | 6 |
| | Images collections | 5 | 5,532 | 16 | 57 | 1 | 2 | 4 |
| | Interactive tools | 11 | 10 | 3 | 77 | 6 | 2 | 4 |
| | Job information | 2 | 188 | 0 | 0 | 4 | 0 | 0 |
| Journals | | 22 | 519 | 4 | 25 | 2 | 4 | 5 |
| Organization | Academy | 4 | 290 | 9 | 27 | 1 | 1 | 6 |
| | Alumni | 1 | 230 | 0 | 0 | 0 | 0 | 0 |
| | Club | 1 | NA | 0 | 0 | NA | NA | NA |
| | Consortium | 1 | 48 | 1 | 1 | 0 | 0 | 0 |
| | Cultural Association | 1 | 4 | 3 | 6 | NA | 4 | 3 |
| | Foundation | 1 | 23 | 0 | 0 | NA | 0 | 0 |
| | Health institution | 1 | 2 | 1 | 2 | 3 | 0 | 3 |
| | High School | 1 | NA | 0 | 0 | NA | NA | NA |
| | Institute | 1 | NA | 0 | 0 | NA | NA | NA |
| | Library | 1 | 274 | 1 | 98 | 2 | 0 | 12 |
| | Master Degree | 1 | 2 | 0 | 0 | 1 | 0 | 0 |
| | National Research Foundation | 1 | 6 | 5 | 13 | 1 | 6 | 1 |
| | Research Center | 27 | 3,835 | 64 | 5,073 | 2 | 7 | 10 |
| | Research Community | 37 | 33,565 | 45 | 20,762 | 6 | 7 | 11 |
| | Research Council | 1 | 72,207 | 647 | 512,869 | 0 | 40 | 37 |
| | Research Federation | 1 | 4,447 | 370 | 12,831 | 7 | 41 | 36 |
| | Research Group | 133 | 3,008 | 12 | 3,692 | 4 | 4 | 6 |
| | Research Institute | 5 | 28,067 | 27 | 9,872 | 25 | 5 | 12 |
| | Research Society | 8 | 140 | 4 | 14 | 3 | 2 | 4 |
| | School | 3 | 45 | 3 | 16 | 3 | 8 | 4 |
| | Scientific Academy | 5 | 5,248 | 22 | 151 | 4 | 5 | 10 |
| | Scientific Association | 13 | 155 | 21 | 399 | 3 | 7 | 7 |
| | Social Group | 1 | 170 | 9 | 14 | 30 | 8 | 13 |
| | Standardization Organization | 1 | 30 | 8 | 160 | 7 | 8 | 12 |
| | University | 2 | 86 | 6 | 6 | 0 | 0 | 6 |
| | University Department | 1 | 2,324 | 1 | 18 | 30 | 11 | 1 |
| | Working group | 1 | 148 | 2 | 515 | 0 | 15 | 9 |
| Personal Websites | | 324 | 98 | 4 | 30 | 7 | 2 | 4 |
| Portal | | 13 | 2,276 | 18 | 49 | 7 | 0 | 7 |
| Raw data | | 15 | 928 | 3 | 137 | 14 | 1 | 5 |
| Research Platform | | 3 | 18 | 5 | 25 | 1 | 2 | 7 |
| Research Project | | 49 | 2,048 | 7 | 63 | 5 | 4 | 8 |

Avg UI: average URLs indexed; Avg RD: average Referral Domains; Avg EBL: average External Backlinks; Avg EOL: average External Outlinks; avg TF: average *Trust Flow*; Avg CF: average *Citation Flow*.
NA: Data no available.

Figure 6 includes a detailed webometrics profile for three specific scholarly websites (research groups, research projects, and academic personal websites). Results evidence the general low impact of websites belonging to these categories as well as a low coefficient of determination between *Trust Flow* and *Citation Flow* on one hand, and between the external backlinks and URLs indexed on the other, variables which use to correlate to each other.

The correlation between *Trust Flow* and *Citation Flow* is slightly higher for research groups websites ($R^2$= 0.45). Moreover, websites included in this category exhibit higher number of indexed URLs and external backlinks, while personal websites and research projects achieve lower web impact.

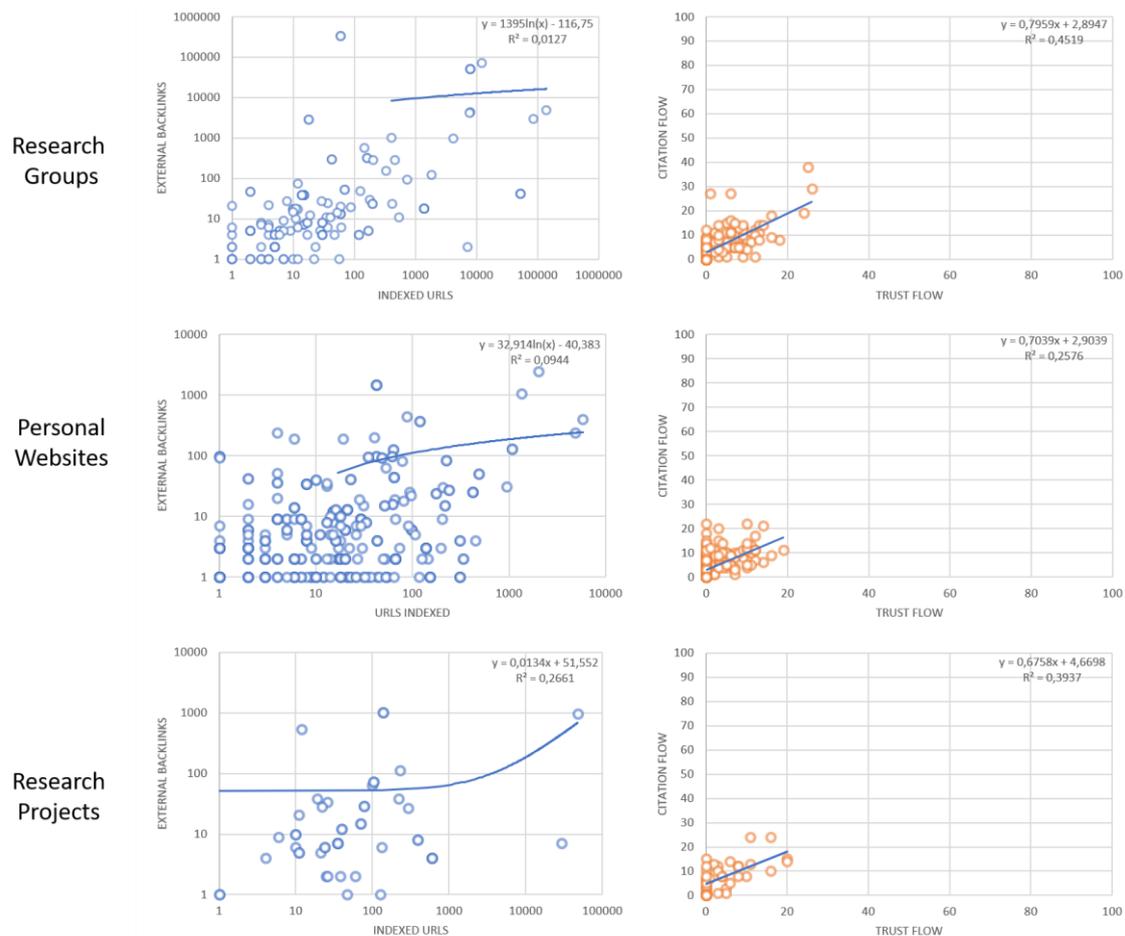

**Figure 6**
Scatterplot of webometric indicators: External backlinks vs ULRs indexed (left) and *Citation Flow* vs *Trust Flow* (right) for different scholarly website categories: (up) Research groups (n= 133); (middle) personal websites (n= 324); (bottom) Research projects (n= 49).

Average values for website typologies inform about high content volumes for databases, and web impact for research communities. However, these results are biased by few outstanding websites and lots of minor websites; therefore impact data at the category-level should be taken cautiously.

In order to filter the most outstanding academic websites regardless their web typology, Table 11 includes the top websites according to the quantity of online contents generated (number of URLs indexed). As we can see, websites devoted to bibliographic and data services (databases, directories, code) and both institutional (research centers) and personal websites are those with greater volume of data generated. Otherwise, the *International Science Council* (council-science) holds the website with the highest number of backlinks received, along with a high web reputation. Notwithstanding, it is remarkable the existence of six websites (among the top 25 greater websites) with less than 100 backlinks received.

**Table 11**
Top 25 primary dot-science scholarly websites according to volume of contents generated

| Web domain | Category | Subcategory | UI | RD | EBL | RD EDU | EBL EDU | TF | CF |
|---|---|---|---|---|---|---|---|---|---|
| communitydata.science | Organization | Research Community | 946,840 | 574 | 15,851 | 3 | 4 | 21 | 28 |
| moonbutt.science | General Content | Code | 637,766 | 103 | 13,472 | 0 | 0 | 7 | 29 |
| proceedings.science | Academic Content | Database | 436,569 | 120 | 34,717 | 9 | 12 | 13 | 49 |
| alephzero.science | Academic Content | Directory | 369,911 | 0 | 0 | 0 | 0 | 0 | 0 |
| fias.science | Organization | Research Institute | 139,795 | 121 | 40,962 | 4 | 5 | 14 | 30 |
| brainstem.science | Organization | Research Group | 138,526 | 140 | 4,830 | 1 | 1 | 1 | 27 |
| lehrbuch.science | Academic Content | Database | 101,904 | 10 | 21 | 0 | 0 | 0 | 9 |
| nes.science | Thematic Website | Technical information | 100,056 | 11 | 500 | 0 | 0 | 4 | 14 |
| lhs.science | Organization | Research Community | 90,773 | 1 | 6 | 0 | 0 | 0 | 5 |
| elvis.science | Organization | Research Group | 84,939 | 7 | 2,997 | 0 | 0 | 5 | 11 |
| council.science | Organization | Research Council | 72,207 | 647 | 512,869 | 44 | 1,079 | 40 | 37 |
| eten.science | Thematic blog | Uncategorized | 70,550 | 6 | 11,757 | 0 | 0 | 14 | 15 |
| astrosophy.science | Organization | Research Center | 65,159 | 18 | 4,128 | 0 | 0 | 11 | 13 |
| tcqp.science | Organization | Research Group | 51,952 | 11 | 42 | 1 | 1 | 13 | 8 |
| psiram.science | Thematic Website | Esoterism information | 50,911 | 47 | 205 | 0 | 0 | 8 | 4 |
| credo.science | Research Project | Uncategorized | 47,797 | 125 | 957 | 5 | 87 | 11 | 24 |
| nouvelle-aquitaine.science | Organization | Research Community | 47,009 | 185 | 395,626 | 0 | 0 | 28 | 34 |
| nova-system.science | Research Project | Uncategorized | 29,137 | 1 | 7 | 0 | 0 | 0 | 4 |
| vshm.science | Organization | Scientific Academy | 24,724 | 49 | 161 | 0 | 0 | 12 | 19 |
| rhyme.science | Academic Content | Database | 23,088 | 23 | 4,879 | 1 | 10 | 0 | 21 |
| cadabra.science | General Content | Application | 21,153 | 100 | 2,850 | 5 | 15 | 16 | 28 |
| isidore.science | Academic Content | Database | 17,413 | 566 | 429,118 | 13 | 29 | 22 | 35 |
| evolutionnews.science | Portal | Uncategorized | 16,884 | 19 | 42 | 0 | 0 | 0 | 7 |
| crest.science | Organization | Research Center | 13,127 | 353 | 66,403 | 15 | 1,367 | 25 | 43 |
| experimenta.science | Organization | Research Center | 12,776 | 706 | 29,184 | 3 | 4 | 24 | 36 |
| mirandalab.science | Organization | Research Group | 12,087 | 311 | 71,359 | 1 | 2 | 0 | 0 |
| badal.science | Thematic Website | Domestic animals | 11,746 | 7 | 314 | 0 | 0 | 0 | 0 |

UI: URLs indexed; RD: Referral Domains; EBL: External Backlinks; RD EDU: Educational Referral Domains; EBL EDU: Educational External Backlinks; TF: *Trust Flow*; CF: *Citation Flow*.

## 5. Discussion

An analysis of 13,900 web domains registered with the dot-science gTLD has been carried out, including measures of web domain's activity and websites' category, volume, and impact. Considering the results obtained, the following aspects should be discussed.

First, ICANN provides top-level domain activity data on a monthly basis. Therefore, data for shorter periods (daily or weekly) are not available. However, the monthly activity is considered sufficient to show the TLD usage over time. Otherwise, activity data on individual web domains were not available due to the impossibility of accessing log files or analytic tools of each particular website. Applications offering web traffic data by means of users' panels, such as *Alexa* or *Similar Web*, were finally discarded due to their lower coverage outside United States, especially China, which holds the 20% of all the dot-science gTLDs registered.

Second, the taxonomy of web domains designed, as all classifications, has an intrinsic subjective component. Categories and subcategories were difficult to establish, especially for non-academic websites. In addition to this, some categories needed special manual review to be clearly distinguished. For example, some parked websites used to include an 'under construction' message just to prevent the appearance of '404 Error - Page Not Found', and others just included a default text, making it difficult to distinguish these websites from other incomplete or flat pages. Therefore, the classification process was slow. The 17% of disagreement achieved in the inter-coder reliability test reflects the difficulty of this categorization process. However, the effects of this disagreement on the final results are marginal, because disagreements were produced mainly by websites that had changed their status (becoming defective or parked websites) without web impact data. The results also found many 'connections refused'. Sometimes this is due to servers not accepting requests from certain geographical locations (or TLDs). It could be useful to test them with anonymous proxies in future studies. Despite these difficulties, the final taxonomy not only represents reliably the type of websites analyzed, but also is exhaustive enough to be extrapolated to other similar works, and constitutes one of the contributions of this work.

Third, web data is subject to high variability. The number of web domains varies from week to week. A website may keep active today and be expired the following week, an access error may be fixed, a webpage may massively get/lose backlinks or new influential websites can be created. Results should be taken into account as a picture of what all web domains registered with the dot-science gTLD look like in the period analyzed. In this sense, the picture obtained is accurate enough to answer the established research questions.

Fourth, Majestic was used as a link provider. Therefore, the accuracy of the web impact data obtained depends on the potential biases that the algorithms used by Majestic both to crawl the World Wide Web and build the tailored indicators – especially Trust Flow and Topical Trust Flow) might introduce. These are proprietary metrics whose complete algorithm is unknown, though their basic functionalities are publicly available as a patented system for the categorization of interlinked information items (Chudnovskiy 2017; 2019).

As regards the primary academic-related dot-science web domains identified, results highlight the presence in absolute numbers of personal websites, research groups and blogs, categories previously analyzed under webometric techniques.

The information structure and contents provided by academic personal websites (main achievements, publication lists, job experience, skills, etc.) are similar to those identified by Mas-Bleda and Aguillo (2013), who analyzed highly-cited researchers' personal websites. Likewise, a significant amount of authors belong to Computer science and tech-related disciplines, an aspect previously commented on by the literature (Mas-Bleda and Aguillo 2013; Barjak 2006). Despite samples are not comparable at all, the low volumes of backlinks and outlinks have been detected in all studies.

Research group websites have been previously studied in different contexts, under institutional (Orduna-Malea 2013) and thematic (Thelwall et al. 2008) filters. These studies highlight the average low quantity of backlinks received. While these results cannot be directly compared with those obtained in this study (web domains under dot-science gTLD), patterns about their low web impact arise.

In the case of blogs, it should be clarified that most of blogs identified cannot be strictly classified as research blogs (Luzón 2009; Shema, Bar-Ilan, and Thelwall 2012), but blogs created by experts. Most of them act as personal websites in blog format and posts are necessarily not intended to disseminate or discuss academic content. For this reason this category should remain differentiated in table 9.

Content analysis for each blog (and any other website under dot-science gTLD) is out of the scope of this study, centered in identifying the volume, category and impact of all web domains registered under the dot-science gTLD. Future works dealing with specific uses and purposes at the content-level are therefore advisable.

## 6. Conclusions

*RQ1: Dot-science gTLD usage*

In response to the RQ1, it is concluded that the dot-science gTLD is not widely used. A total of 13,900 web domains were found worldwide. That is, the user demand for this gTLD since its inception has been low. In addition to this, an important drop in the number of both DNS and *whois* queries were detected over time, evidencing a lack of interest both from registrars and registrants. Furthermore, a significant percentage of the web domains were registered by few registrar companies, mainly for cybersquatting or reselling purposes.

*RQ2: Dot-science gTLD types*

In response to the RQ2, a significant percentage (35.5%) of all web domains registered with the dot-science gTLD corresponded to parked web domains. Another important percentage (38.7%) of web domains experienced some access error. This circumstance is partially attributed to parked web domains hosted with bad quality hosting services and low technical maintenance. Most of these web domains end up expiring or generating permanent access errors. Only 2,222 (16%) web domains had a primary function. However, most web domains were empty websites, websites dedicated to non-academic purposes, or even exhibiting dubious or fraudulent content. Presumably, the dot-science gTLD was used to acquire some false authority or better search engine positioning.

As regards primary scholarly-related dot-science gTLD categories, a total of 1,175 (8.5% of all web domains registered) were found. Considering that the analysis has been performed at the TLD-level, this result confirms a low scholarly usage of the dot-science gTLD.

Among the main web genres, personal academic websites were the most frequently used category (342 web domains). Other popular dot-science website usages were thematic websites (either blog or static pages format), research groups and research centers. The use of these web genres might be due to academics, scholars or independent researchers who cannot use institutional websites, or scholars who prefer to freely create a personal/lab website with an academic branded name to promote their contents and activities instead of using the institutional services (which might be technically constrained or under certain usage policies). Private company websites was another important category. However, most companies were not professionally linked to the scientific activity at all (companies selling Viagra, offering SEO services or spiritual help were found).

*RQ3: Dot-science gTLD web impact*

In response to the RQ3, the number of websites generating large volumes of content is scarce and related to websites with dubious practices. Those websites receiving higher number of backlinks were generally non-academic websites applying intensive link building SEO practices. In this sense, *Trust Flow* has been proved to be an effective method to filter academic content. Few databases and institutional websites (those related to research groups, centers, institutes and communities, and a research council) were found as the most outstanding scholarly websites registered with a dot-science gTLD.

The general low quantity of academic-related websites found could be explained by the fact that currently the online scientific world is mainly shaped by universities, research associations and publishing companies, already established on the Web and with stable TLDs registered, both educative (dot-edu), geographical (dot-es, dot-ac.uk, etc.), commercial (dot-com) and even specific (dot-ai; dot-info). In this sense, a new complementary and specialized gTLD oriented towards academic purposes should appear for webmasters as a reputable place, with flagship institutions, products, or services as touchstone websites, and with marketing actions especially tailored for academics and researchers. However, results reveal dot-science as a low-quality gTLD (i.e., websites with connection errors, high number of parked websites and dubious content). Moreover, registrar pages showed very cheap prices (which makes it appealing for search engine optimization professionals to use these gTLDs for promoting third-party websites), etc. These issues might have compromised the reputation of the dot-science gTLD in the research community, and consequently, its adoption.

These results show two main implications. First, dot-science gTLD can be considered a lost opportunity to have a reputable academic web domain gathering different research projects and initiatives, as well as a communication area for individuals who cannot use their institutional web domains to store and disseminate particular projects related to Science. Second, all dubious content found reveals bad practices on the Web where the tag 'science' is generally used as a mechanism to deceive search engine algorithms.

As a conclusion, and given the results obtained, the feasibility and advisability of transforming the dot-science from a generic to a sponsored top-level domain should be explored. Academic institutions, research societies or non-governmental organizations (such as UNESCO) should be considered to manage a top-level domain aimed at gathering and making accessible to users academic content worldwide.